\documentstyle[epsf,11pt]{article}
\newcommand{\BS}{{{\it Beppo}SAX}\ \ignorespaces}
\newcommand{\ltsima} {$\; \buildrel < \over \sim \;$}
\newcommand{\gtsima} {$\; \buildrel > \over \sim \;$}
\newcommand{\lta} {\lower.5ex\hbox{\ltsima}}
\newcommand{\gta} {\lower.5ex\hbox{\gtsima}}
\begin{document}
\vspace{1.0cm}
{\Large \bf PROGRESS IN UNDERSTANDING BLAZARS FROM \BS OBSERVATIONS}

\vspace{1.0cm}

L. Maraschi$^1$, L. Chiappetti$^2$, G. Fossati$^3$, E. Pian$^4$ and F. 
Tavecchio$^1$

\vspace{0.9cm}
$^1${\it Astronomical Observatory of Brera, via Brera 28, 20121 Milano, 
Italy}\\
$^2${\it~IFCTR, CNR,via Bassini 15, 20133 Milano, Italy}\\
$^3${\it~SISSA/ISAS, Via Beirut 2-4, I-34014 Miramare (Trieste),
Italy}\\
$^4${\it~ITESRE, C.N.R., Via Gobetti 101, I-40129 Bologna, Italy}\\
%$^5${\it~}\\
%$^6${\it~}\\
%$^7${\it~}\\

\vspace{0.5cm}

%%%%%%%%%%%%%%%%%%%%
\section*{ABSTRACT}
%%%%%%%%%%%%%%%%%%%%
Results obtained with \BS observations of blazars 
within various collaborative programs are presented. 
The spectral similarity ``paradigm", whereby the 
spectral energy distributions of blazars follow a sequence,
 leading to a unified view of the whole population,
 is briefly illustrated. We concentrate on recent observations of flares
and associated spectral variability  
% A definite pattern of variability follows  
%whereby corresponding frequencies should vary in a correlated fashion.
%In the case of "red" blazars this has been verified for 3C 279, at least on 
long
%timescales. 
for three objects at the ``blue" end of the spectral sequence,
 namely PKS 2155--304, Mkn 421 and Mkn 501. The results are discussed
in terms of a general analytic synchrotron self-Compton  interpretation
of the overall spectrum. The physical parameters of the quasi-stationary 
emission region can be derived with some confidence, while the
variability mechanism(s) must be complex.

%%%%%%%%%%%%%%%%%%%%%%%
\section{INTRODUCTION}
%%%%%%%%%%%%%%%%%%%%%%%

One of the most poorly understood phenomena in AGN is the origin of 
relativistic jets. Their existence, previously inferred from indirect 
arguments (Blandford \& Rees 1978), was spectacularly demonstrated
by the observation of superluminal expansion of the knots in the radio jets
on the parsec scale. 
The unified scheme of AGN postulates that all radio-loud
AGN possess relativistic jets and that blazars are the subset where the jet
happens to point at a small angle to the line of sight. Because the plasma
flows in the jet at relativistic speed the emitted radiation is concentrated
in a narrow cone (beam) along the direction of motion. An observer at small 
angle 
to the beam will receive radiation  strongly enhanced by Doppler boosting.

In blazars the non thermal continuum received from the jet is largely
dominant over
the more isotropic radiation emitted by the surrounding gas or stars.
Therefore they are the best laboratory to probe the processes at work 
in relativistic jets. Understanding the radiation mechanisms allows 
reconstructing the spectra of relativistic particles and discussing
the mechanisms of particle acceleration and energy transport along the jet  
and ultimately their origin.

%%%%%%%%%%%%%%%%%%%%%%%%%%%%%%%%%%%%%%%%%%%%%%%%%%%%%%%
\section{THE SPECTRAL ENERGY DISTRIBUTIONS OF BLAZARS}
%%%%%%%%%%%%%%%%%%%%%%%%%%%%%%%%%%%%%%%%%%%%%%%%%%%%%%%

Observationally, blazars are characterised by a strong radio core with flat
or inverted spectrum, and by an extremely luminous broad band continuum,
highly
variable at all wavelengths. After the launch of the CGRO it was discovered
that this continuum extends into the $\gamma$-ray range and
most importantly that the $\gamma$-ray luminosity represents a large fraction 
of the total emitted power.     
 
The presence or absence of broad emission lines in the optical-UV spectrum
has led to distinguish quasar-like objects from BL Lac type objects.
However there are arguments to believe that these differences, rather than
representing a genuine dichotomy in the type of processes occurring at the
nucleus
 may arise from different physical conditions outside the jet (e.g.,
 Bicknell 1994). 
Support to this unifying view comes from a comparison of
the overall spectral shapes of different subclasses of blazars. 

To this end we collected multiwavelength data for three complete samples of 
blazars,
the 2 Jy sample of flat spectrum radio-loud quasars (FSRQ) the 1 Jy sample
of BL Lac objects and the X-ray selected BL Lacs from the Einstein Slew Survey
(see Fossati et al. 1998 for full information  and references).
 The results are shown in Fig. 1, where the three samples have been merged and 
the total sample has been binned according to radio-luminosity only.
From the figure it is apparent that:

\begin{itemize} 
\item All the spectral energy distributions (SED) in the $\nu f_{\nu}$ 
representation are
characterized by two peaks, indicating two main spectral components.

\item The first peak falls at lower frequency for higher luminosity objects

\item The second peak frequency seems to correlate with the first one, as 
indicated
by the  $\gamma$-ray and X-ray slopes. The curves drawn as reference correspond
to a fixed ratio between the two peak frequencies.

\end{itemize}

\begin{figure}
\epsfysize=8cm % fix the y-dimension and scales x-dim. to y-dim.
%\epsfxsize=15cm % fix the x-dimension and scales y-dim. to x-dim.
% Feel free to do the choice you prefer but do not exceed the x-dimension
% of the text lines
\vspace*{-0.7cm}
\hspace{-0.2cm}\epsfbox{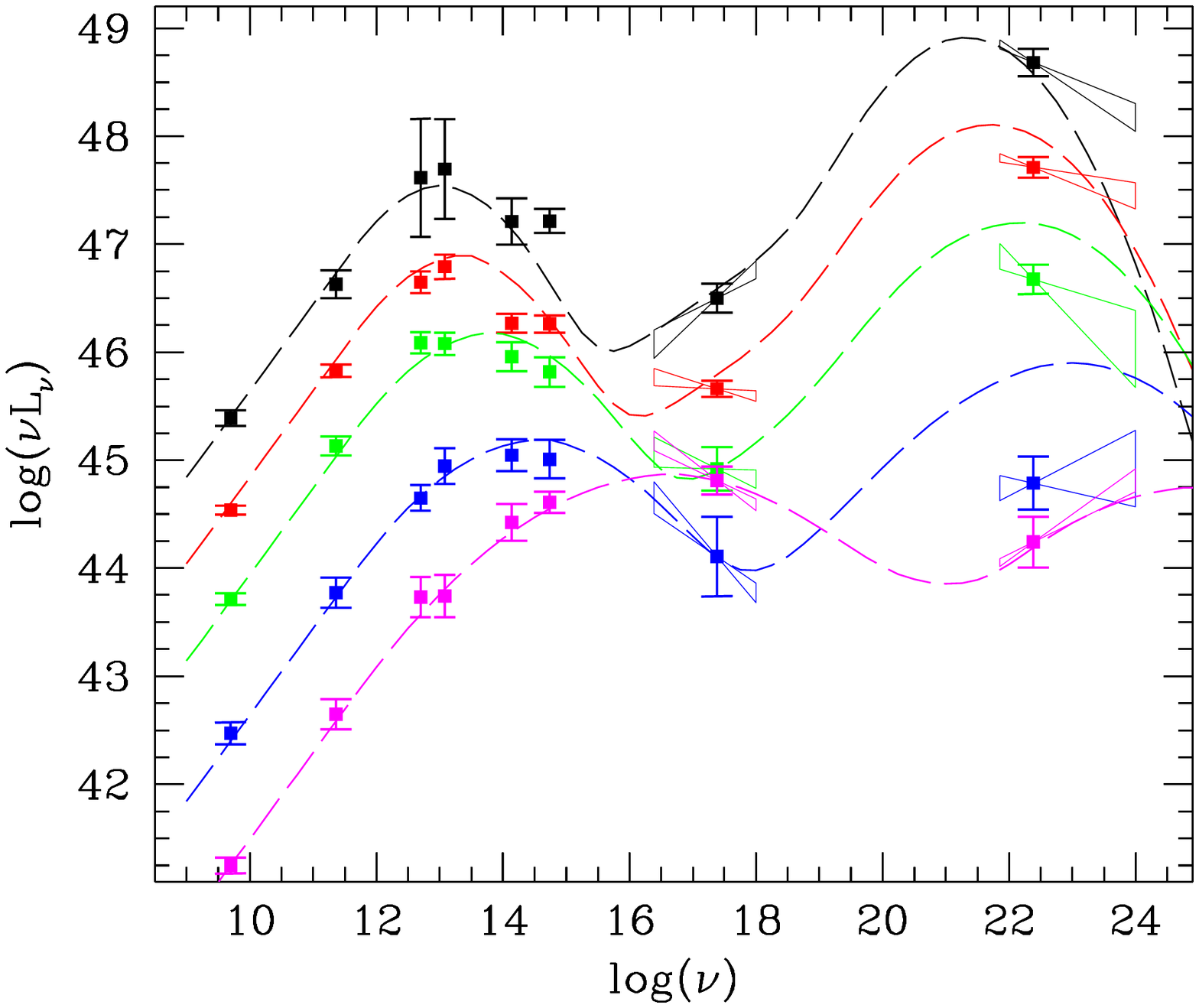}
\epsfysize=8.5cm % fix the y-dimension and scales x-dim. to y-dim.
\vspace{-8.9cm}
\hspace*{8.8cm}\epsfbox{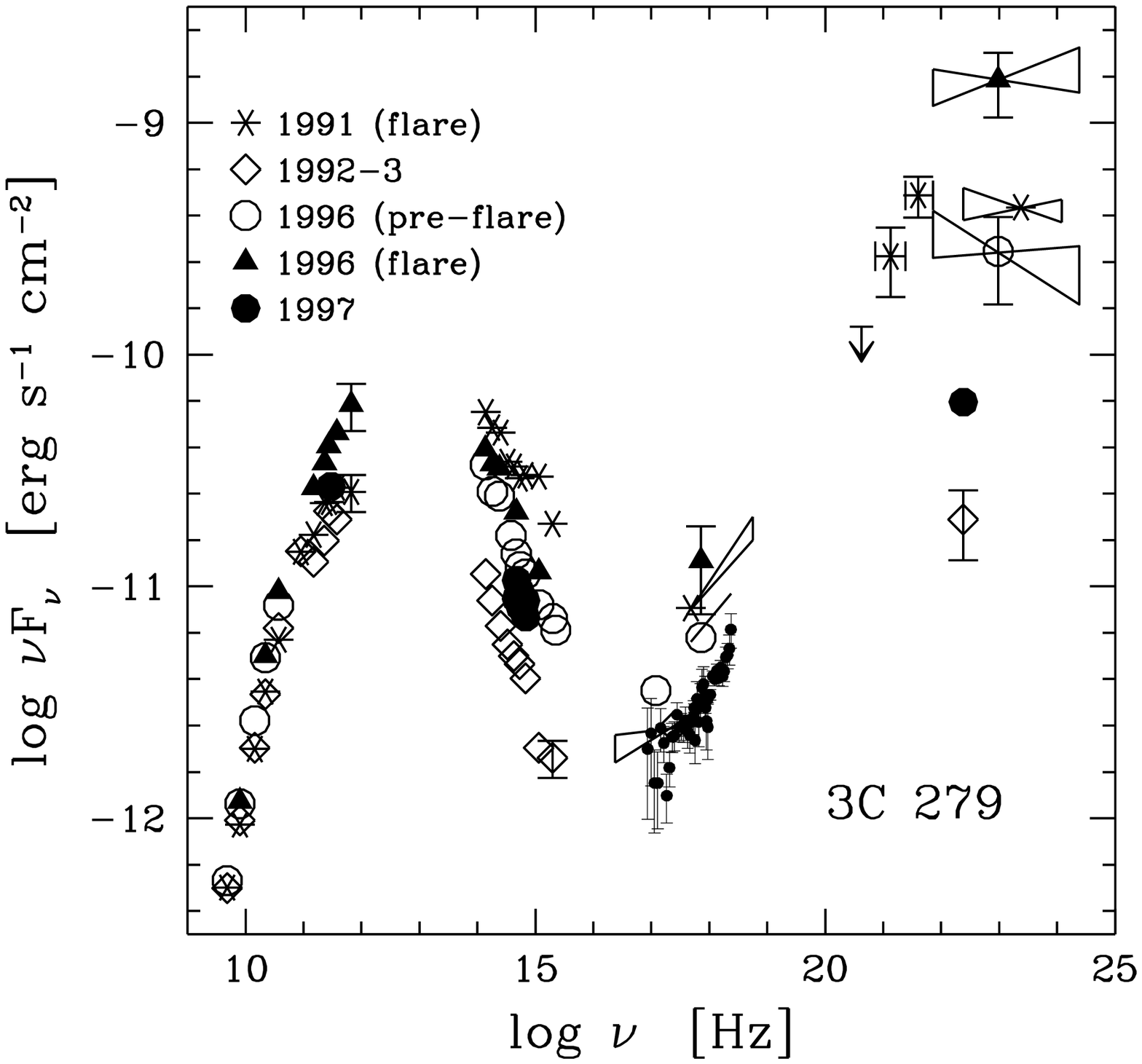}
%for centering: act on hspace argument 
\caption[h]{ {\it Left panel}: Average SEDs of the ``merged" blazar sample
 binned according to radio
luminosity, irrespective of the original classification.  Empty asymmetric
 triangles represent uncertainties in spectral shapes
as measured in the X-ray and $\gamma$-ray bands. The overlayed dashed 
curves are analytic approximations obtained assuming that the ratio of the
two peak frequencies is constant and that the luminosity of the second peak is 
proportional to the radio luminosity (from Fossati et al. 1998).
 {\it Right panel}:  Radio-to-$\gamma$-ray energy distribution of 3C~279 at 
different
epochs. The X-ray spectrum observed with \BS on January 1997 is shown,
 together with the simultaneously measured optical and $\gamma$-ray
fluxes.
 Data obtained in June 1991, and December 1992-January 1993  are from
 Maraschi et al. (1994). The pre-flare
and flaring state data in January-February 1996 are from Wehrle et al.
(1998).
%The slope of the ASCA spectrum ($\alpha_\nu = 0.7$) has been reported
%normalized to the RXTE point closest in time.  
%The EGRET best fit power-law spectra
%referring to the January 16-30 (preflare) and February 4-6 (flare) periods are
%shown, normalized at 0.4 GeV. Errors have been reported
%only when they are bigger than the symbol size.
The UV, optical and near-IR data have been corrected for Galactic 
extinction.

}
\vspace*{.1cm}
\end{figure}

Although one should be aware  that selection biases may affect the results 
(see Fossati et al. 1997), Figure 1 suggests that the SEDs
of all blazars are globally similar and lie along a continuous
spectral sequence.  
For the most luminous objects the first peak falls at frequencies lower
than the optical band, while for the least luminous ones the reverse is true.
Thus highly luminous objects have  steep (``red") optical-UV continua,
while low luminosity objects with peak frequency beyond the UV range
have flat (``blue") optical-UV continua. For brevity we will refer
to objects on the high and low luminosity end of the sequence 
 as ``red" or ``blue" blazars.

%%%%%%%%%%%%%%%%%%%%%%%%%%%%%%%%%%%%%%%%%%%%%%%%%%%%%%%%%
\section{ IMPORTANCE OF MULTIWAVELENGTH VARIABILITY }
%%%%%%%%%%%%%%%%%%%%%%%%%%%%%%%%%%%%%%%%%%%%%%%%%%%%%%%%%
It is generally thought that the first spectral component is due to 
synchrotron radiation. The spectra from the radio to the submillimeter range
most likely involve superposed contributions from different regions of the jet 
with different self-absorption turnovers. From infrared (IR)
frequencies upwards
the synchrotron emission should be thin and could be produced in a single
zone of the jet, allowing adoption of a homogeneous model.

The  second spectral component (peaking in $\gamma$-rays) could be
produced by the high energy electrons responsible for the synchrotron 
component,  upscattering soft photons via the inverse Compton process (IC).
% The seed photons
%for upscattering could be the synchrotron photons themselves 
%(synchrotron self Compton model, SSC) or photons outside the jet
%(external Compton model, EC) possibly produced in an accretion disk
%or torus and scattered or reprocessed by the surrounding gas
(e.g., Sikora 1994; Ulrich, Maraschi, \& Urry, 1997 (UMU97), and references
therein).

An immediate consequence of this interpretation is that changes in the 
electron population should produce correlated variability
in the two components. Different models for the soft photon source(s)
(the synchrotron photons themselves, synchrotron self-Compton, SSC;
photons from a possible accretion
disk, 
or related broad line region, EC; synchrotron photons backscattered from
gas clouds close to the jet, ``mirror Compton", MC, see e.g., UMU97, and
references therein)  
imply different possible ways of variability
of the IC component. For instance in the SSC model
the Compton component must vary with the synchrotron component and with larger
amplitude (in the Thomson regime), while in the EC model variations of the IC 
component 
without associated variations in the synchrotron component should be possible
(Ghisellini \& Maraschi 1996). However in the case MC model
variability could closely mimick the SSC type.
Clearly  the study of correlated variability at frequencies close to the two 
peaks
 of the SED is an essential  tool to constrain models.

It is important to stress that  in blue blazars the X-ray range 
represents the high energy end of the synchrotron component.
Deriving from extremely energetic electrons, with short lifetimes
 the X-ray emission from blue blazars is  rapidly variable. 
The associated IC emission falls beyond the EGRET energy range and is detectable
 in the TeV band for the few brightest objects. This offers the possibility 
of ground based monitoring at the highest energies (VHE) and makes
 the study of the X-ray/TeV correlation extremely interesting.

For red blazars the corresponding energy ranges are from 
the IR to the UV (1-10 eV) for the synchrotron component and  
from the MeV to the GeV bands for the IC component, while X-rays represent
the low energy end of the IC component. Correlation studies of the IR to UV 
emission on one hand with the X-ray to $\gamma$-ray emission on the other
have been performed by several groups (see UMU97 and refs therein).
The most intensively observed object has been 3C 279. Repeated intensive 
multiwavelength campaigns have detected a series of high and low states
with rather regular spectral variations. The synchrotron intensity
is indeed correlated with the IC intensity at least on long time scales
with $\gamma$-rays showing the largest variability amplitude. The \BS
data obtained in 1997, when the source was in an intermediate 
intensity state, confirm this trend (see Fig. 1, from Maraschi 1998).
 
Unfortunately, the exhaustion of gas is causing substantial 
degradation of the EGRET sensitivity, practically preventing further monitoring 
in $\gamma$-rays. At the same time  the developing capabilities of the 
ground based Cherenkov telescopes allow improved monitoring at TeV energies,
shifting the focus of multiwavelength studies  from red to blue blazars. 

Therefore we will concentrate here  on
\BS results concerning three bright
 blue blazars detected in the TeV band, PKS 2155--304, Mkn 501 and Mkn 421.
 Other important results 
 obtained with \BS concern the spectral variability
of the third blue BL Lac detected at TeV energies 
(Giommi et al. 1998a),
 the X-ray spectra of bright, red blazars
(e.g., Ghisellini et al. 1998) and of other BL Lac samples
(e.g., Wolter et al. 1998, Padovani et al. 1998).

%%%%%%%%%%%%%%%%%%%%%%%%%%%%%%%%%%%%
\section{PKS 2155--304}
%%%%%%%%%%%%%%%%%%%%%%%%%%%%%%%%%%%%

PKS 2155--304 is one of the brightest BL Lacertae objects in the X-ray
band and one of the few detected in $\gamma$-rays by the EGRET experiment
 on  CGRO (Vestrand et al. 1995). It was observed by \BS during the PV phase
(Giommi et al. 1998b).
No observations at other wavelengths simultaneous with the $\gamma$-ray ones
were ever obtained for this source, yet it is essential to measure the
IC and synchrotron peaks at the same time in order to constrain
emission models unambiguously 
(e.g., Dermer et al. 1997, Tavecchio, Maraschi, \& Ghisellini 1998).
For these reasons, having been informed by the EGRET team of their
observing plan and of the positive results of the first days of the CGRO
observation, we asked to swap a prescheduled target of our \BS
blazar program with PKS 2155--304. In 1997 November 11-17 (Sreekumar
\& Vestrand 1997) the $\gamma$-ray flux from PKS 2155--304 was very high,
 roughly a factor of three
greater than the previously published value from this object. \BS
pointed PKS 2155--304 for about 1.5 days starting on November 22. Quick-look
analysis indicated that also the X-ray flux was close to the highest
recorded levels  (Chiappetti \& Torroni 1997).
A paper on these data is currently submitted for publication (Chiappetti et al. 
1998).
Here we summarise the most important results.

i) {\it Light Curves}

Fig. 2 (left panel)  shows the light curves  binned over 1000 sec obtained
 in different energy bands, 0.1-1.5 keV (LECS) and 3.5-10 keV (MECS).
 The light curves show a clear high amplitude
variability:  three peaks can be identified.  
The most rapid variation observed (the decline from the peak
at the start of the observation) has a halving timescale 
of about $2\times  10^{4}$ s, similar to previous occasions
(see e.g., Urry et al. 1997). No shorter time scale variability is detected 
although our observations
would have been sensitive to doubling timescales of order $10^{3}$ s.
 The variability amplitude is energy dependent being larger at higher energies.
The hardness ratio correlates positively with the flux, indicating that states 
with
higher flux have harder spectra.

\begin{figure}
\vspace*{-1.0cm}
\epsfysize=8.5cm % fix the y-dimension and scales x-dim. to y-dim.
%\epsfxsize=15cm % fix the x-dimension and scales y-dim. to x-dim.
% Feel free to do the choice you prefer but do not exceed the x-dimension
% of the text lines
\hspace{-0.1cm}\epsfbox{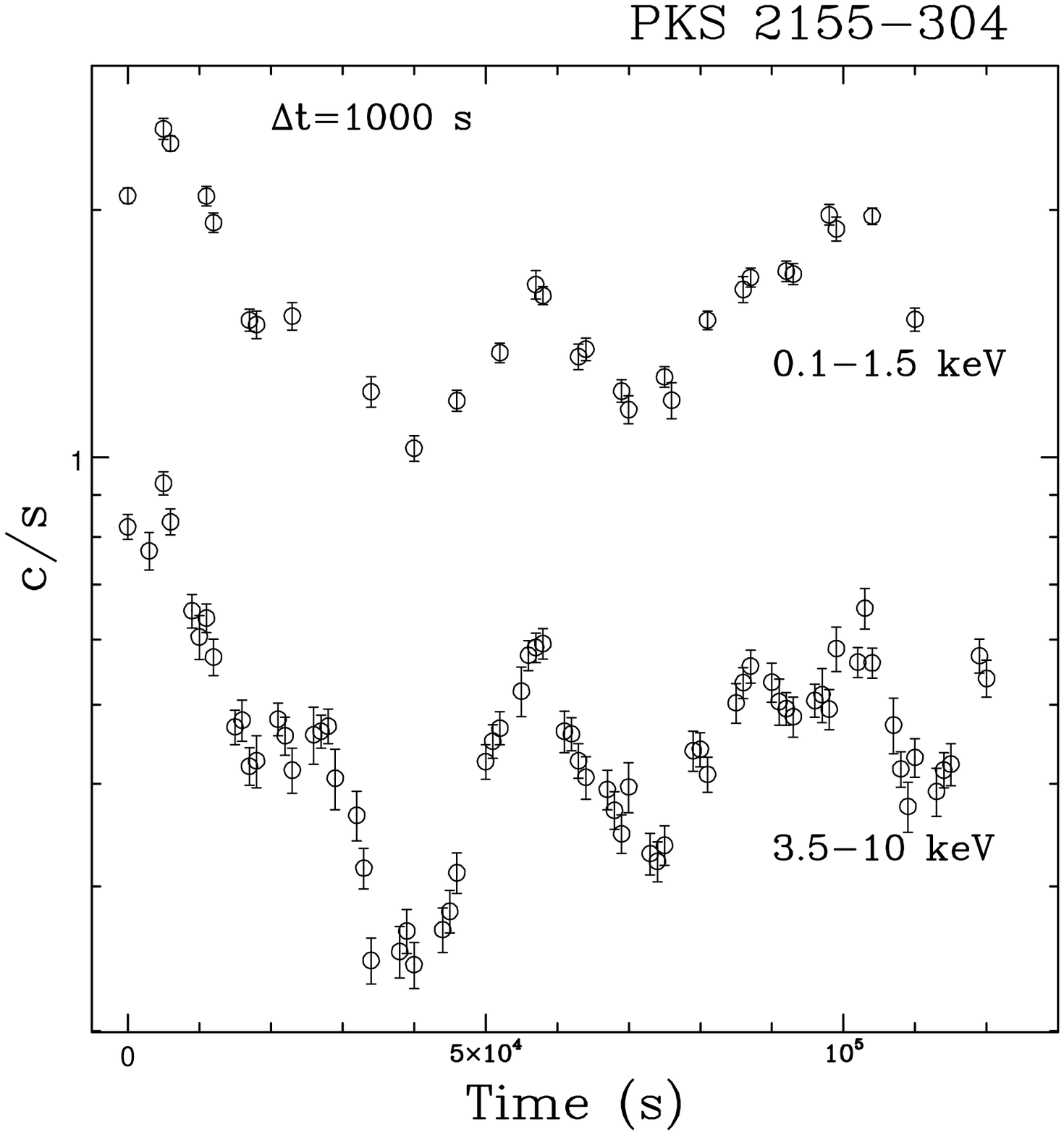}
\epsfysize=9.3cm % fix the y-dimension and scales x-dim. to y-dim.
\vspace{-8.75cm}
\hspace*{8.1cm}\epsfbox{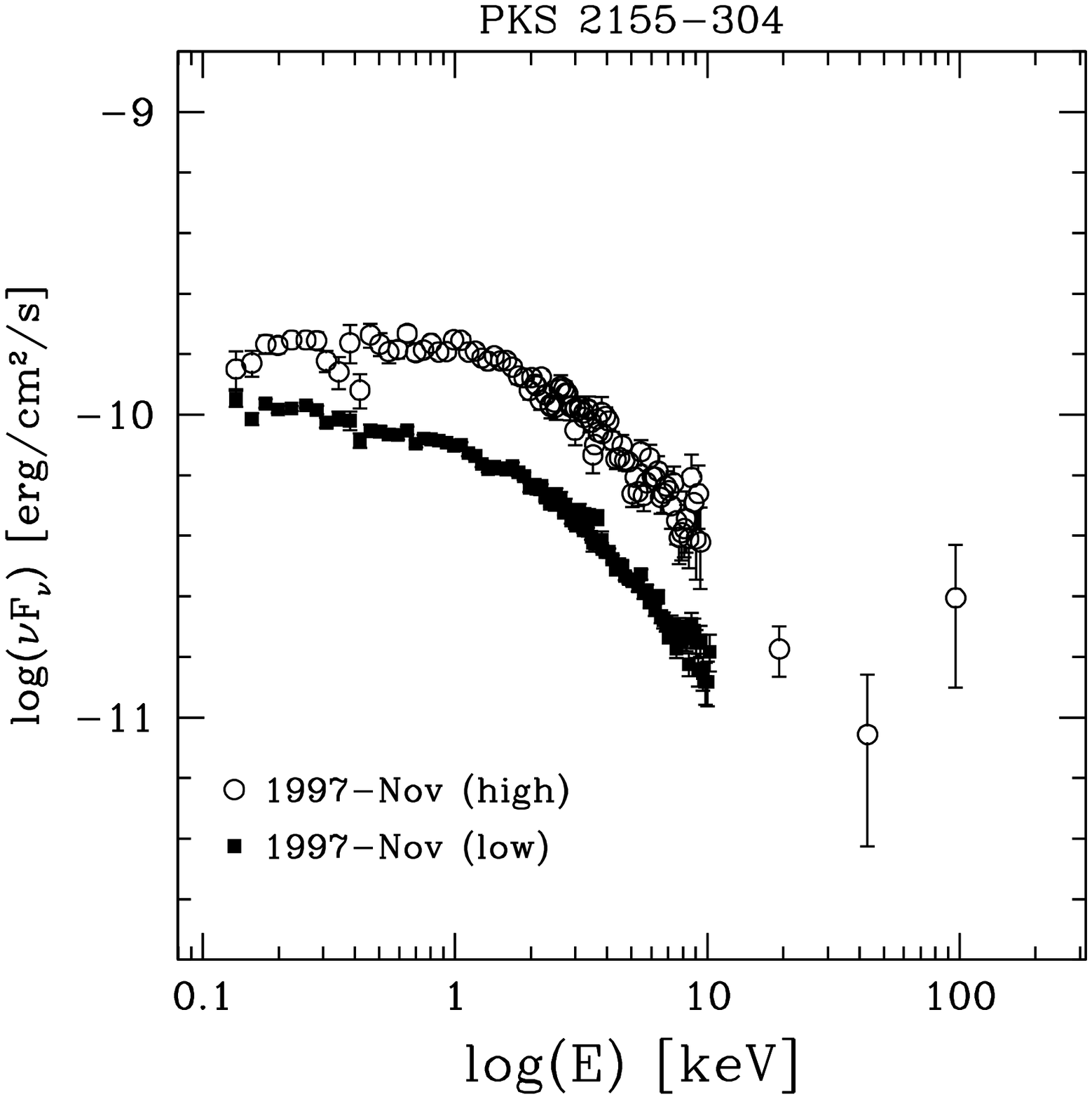}
%for centering: act on hspace argument 
\caption[h]{{\it Left panel}: Soft (0.1--1.5 keV) and medium (3.5--10 keV)
light curves (in logarithmic scale) obtained from the November 1997 observation 
of PKS 2155--304 with \BS. 
{\it Right panel}: Deconvolved X-ray
 spectra of PKS 2155--304 in the highest and lowest states observed in November
1997. The open circles between 10 and 100 keV represent the PDS data, which are
averages over the whole observation.
}
\end{figure}
We looked for time lags between variations at different energies as suggested
by previous ASCA observations of the same source and of  Mkn 421 (Makino
this volume; Takahashi et al. 1996). The possible presence of a soft lag
is indicated by the fact that the maximum hardness ratio occurs {\it before}
maximum intensity. Using the Discrete Correlation Function (Edelson \& Krolik 
1988) and
fitting its maximum
with a Gaussian we estimate from its peak a lag of $0.49 \pm 0.08$ (1-$\sigma$).
A more detailed  discussion of the issue of lags in this source, including
comparison with previous observations and error estimates through
Monte Carlo simulations is given in Treves et al. (1998).

ii) {\it Spectra}

We found that the {\it joint} LECS and MECS spectra could not be adequately 
fitted
by either a single or a broken power law (bpl). A single power law is 
unacceptable
even in  each spectral range while  a bpl model with Galactic absorption
yields good fits to the data of each instrument with consistent slopes at the 
high
energy end of
the LECS and at the low energy end of the MECS. 
We therefore adopted this model as an approximation to a more realistic
continuously curved spectral shape (see Giommi et al. 1998b)
The spectral indices  derived from separate bpl fits to the LECS and MECS
data integrated over the whole observation are
reported in Table 1, which allows comparison with the other two sources 
discussed
below.
The change in slope between the softest (0.1-1 keV) and hardest
 (3-10 keV) energy ranges is $\simeq 0.8$.

Fitting together the  MECS and PDS data yields  spectral parameters 
very similar to those  obtained for the MECS alone. The residuals show
that the PDS data are 
consistent with an extrapolation of the MECS fits up to about 50 keV.
Above this energy the PDS data show an excess indicating a
flattening of the spectrum.

The spectrum at the flare peak is harder than at lower intensity, as can be seen 
by
computing directly the ratio of the count rate spectra as a function of energy, 
yet the
spectral change is small and for bpl fits of the separate instruments
the derived parameters are only marginally different. The deconvolved spectra 
are
shown in Fig. 2 (right panel).

%%%%%%%%%%%%%%%%%%%%%%%%%%%%%%%%%%%%
\section{Mkn 421}
%%%%%%%%%%%%%%%%%%%%%%%%%%%%%%%%%%%%

Mkn 421, closely similar to PKS 2155--304 in brightness and spectral shape at UV 
and
X-ray wavelengths was observed by \BS in April 1998  as part of a large
multiwavelength campaign based on a week of continuous observation with ASCA and
simultaneous monitoring  with the available Cherenkov telescopes, Whipple,
HEGRA and CAT. The \BS observations were scheduled before the
ASCA ones
in order to extend the coverage in time. 
A well defined pronounced flare was observed at the beginning of the observation
reaching the peak in about 3 hrs and decaying in half a day. Due to its bright 
state
the source was well visible also in the high energy instrument, the PDS.

i) {\it Light Curves}

The light curves in three energy bands, derived from the LECS, MECS and PDS,
all normalised  to their respective 
 average intensity, are shown in Fig. 3. The amplitude of
variability increases  and the decay time scales decrease 
with increasing energy although the latter effect is difficult to quantify 
without a
specific model for the light curve.

\begin{figure}
\vspace*{-0.7cm}
\epsfysize=9cm % fix the y-dimension and scales x-dim. to y-dim.
%\epsfxsize=15cm % fix the x-dimension and scales y-dim. to x-dim.
% Feel free to do the choice you prefer but do not exceed the x-dimension
% of the text lines
\hspace{0.9cm}\epsfbox{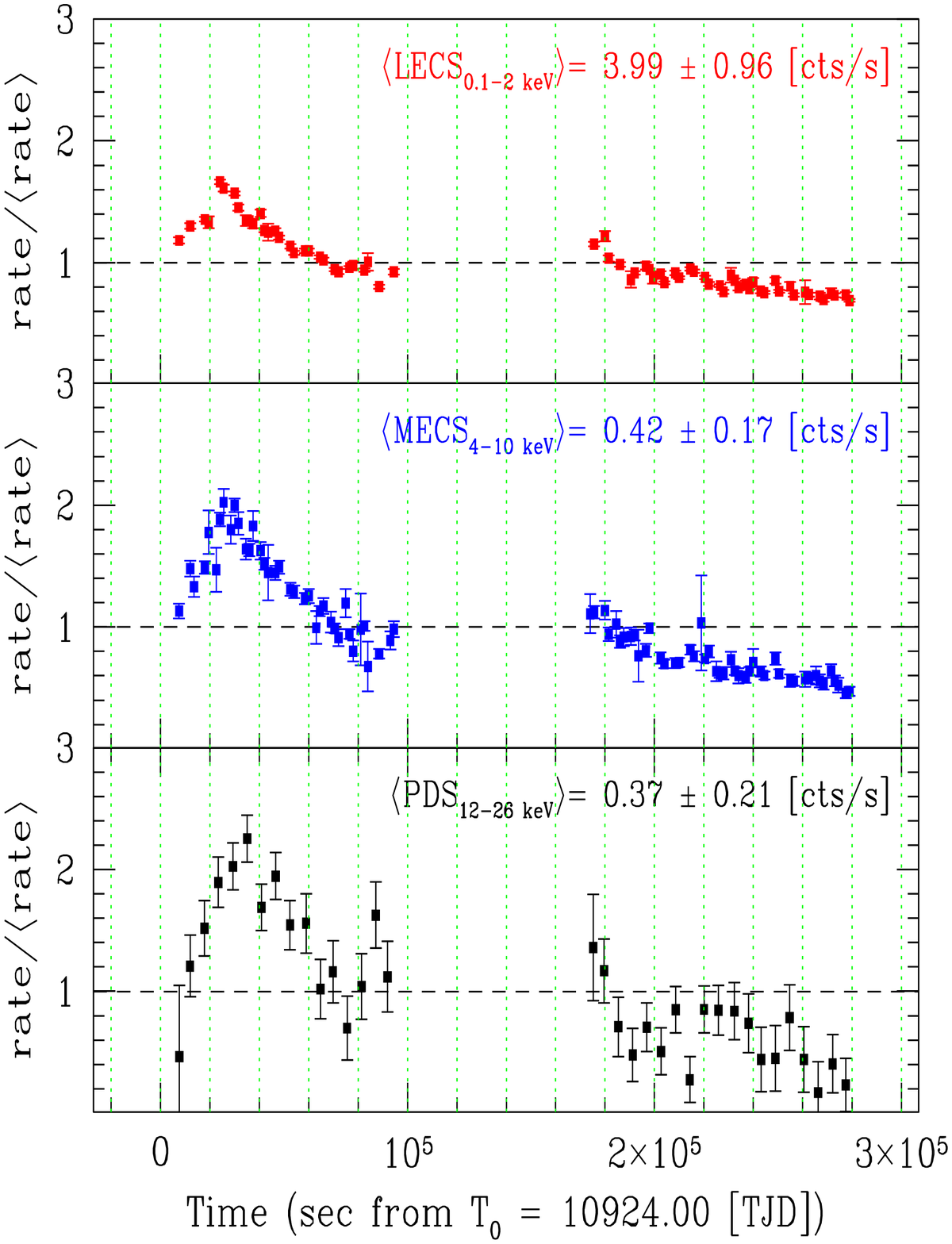}
\epsfysize=8.5cm % fix the y-dimension and scales x-dim. to y-dim.
\vspace{-8.75cm}
\hspace*{8.0cm}\epsfbox{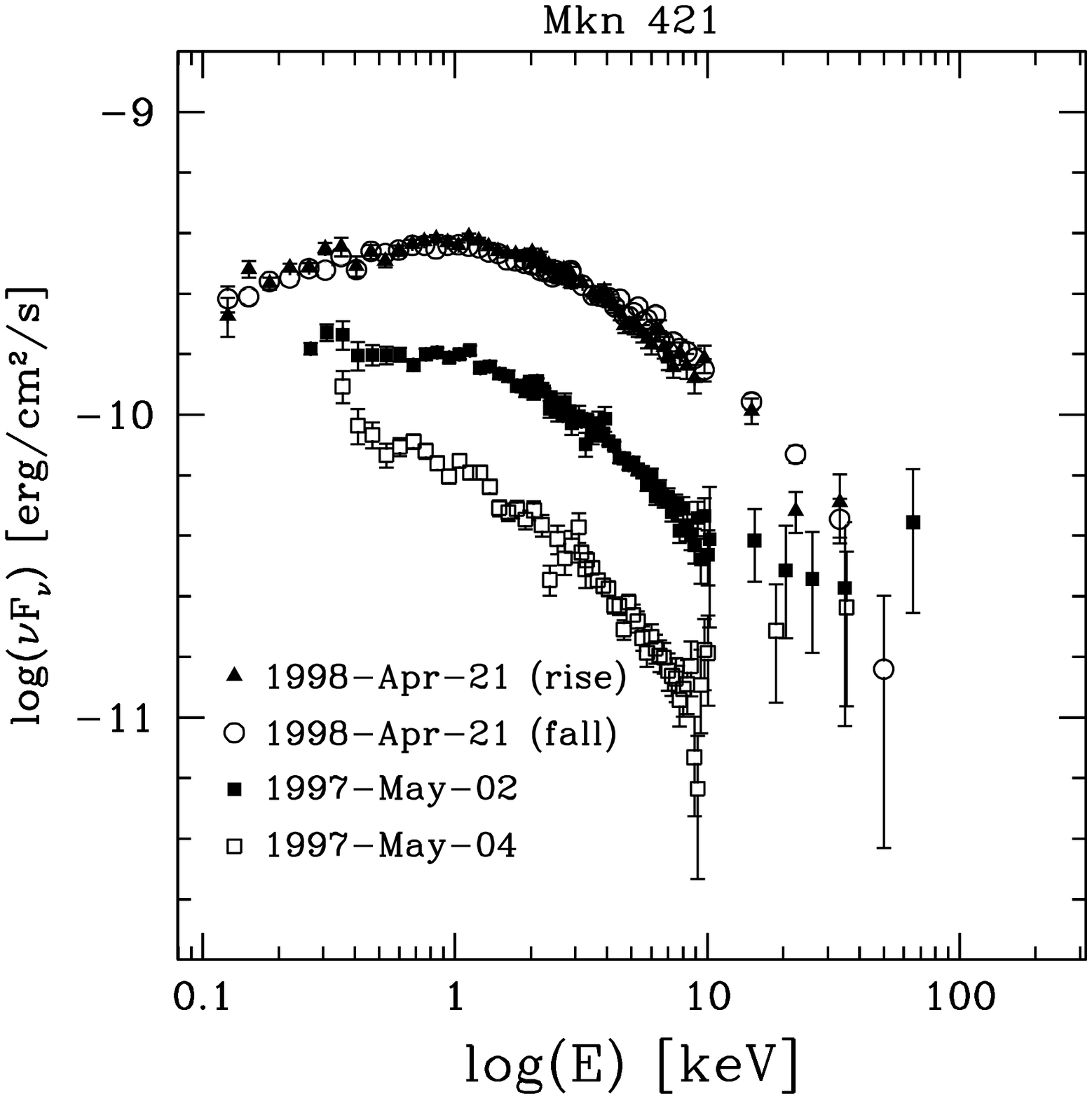}
%for centering: act on hspace argument 
\caption[h]{{\it Left panel}: light curves of Mkn 421
observed in April 21-24 in the
0.1-2 keV  (LECS), 4-10 keV (MECS) and 12-26 keV (PDS) energy bands 
(top to bottom). Average count rates in each band with their dispersions are
given.
{\it Right panel}: Deconvolved X-ray spectra obtained with \BS 
in 1998 and 1997.
}
\end{figure}

It is interesting to explore whether the light curves at different frequencies 
show 
lags as observed previously in this source by Takahashi et al. (1996).
A first analysis with the DCF method reveals that, contrary to the case of 
PKS 2155--304 and of the 1994 flare of Mkn 421, in the present flare
 the soft photons  {\it lead} the medium energy ones by about 1500 sec.
The significance of this result needs to be assessed by a reliable estimate
of the errors of the DCF method, however we can definitely exclude that lags or 
leads
larger than this value are present. 

ii){\it Spectra}

The shape of the X-ray spectra is qualitatively similar to the case of PKS 
2155--304
and similar considerations apply. The spectral indices  derived from separate 
bpl fits to the LECS and MECS data for 1998 April 21, 
which cover the whole flare, and for 1997 May 4 are
reported in Table 1.
The change in slope between the softest (0.1-1 keV) and hardest
 (3-10 keV) energy ranges is $\simeq 0.8$ in both cases.
Deconvolved spectra are compared in Fig. 3. It is apparent that the peak in
the power distribution moves to higher energies with increasing intensity,
reaching 1 keV during the flare.

%%%%%%%%%%%%%%%%%%%%%%%%%%%%%%%%%%%%
\section{Mkn 501}
%%%%%%%%%%%%%%%%%%%%%%%%%%%%%%%%%%%%

\BS observations of Mkn~501 in April 1997 revealed a completely new
behavior. The spectra showed that at that epoch the
synchrotron component peaked at 100 keV or higher energies, implying a shift
of at least two orders of magnitude of the peak energy with respect to the
quiescent state (Pian et al. 1998). 
 Correspondingly the source was extremely bright in the
TeV band and exhibited rapid flares (Catanese et al. 1997, Aharonian et al. 
1997). 
The source was reobserved with \BS
 at three epochs, on 28, 29 April
and 1 May 1998, for $\sim$10 hours, simultaneously with ground-based
optical and TeV Cherenkov telescopes (Whipple and HEGRA).

  For all epochs, fits to all the data
either with a single or a broken power-law are unacceptable. The ``curvature" 
in the LECS-MECS range is however smaller than for the previous two sources
and a bpl fit to the joint LECS-MECS spectra gives a satisfactory result. 
The joint MECS, HPGSPC and PDS spectra are also well fitted with a bpl model.
The spectral indices, obtained by fixing the
value of the hydrogen column density to the Galactic value ($1.73 \times
10^{20}$ cm$^{-2}$, from Lockman \& Savage 1995), are reported in Table~1. 
The deconvolved spectra
of 1998 are compared in Fig. 4 (right panel) with that of the 1997 flare. 

\begin{table}
\caption{Comparison of spectral parameters of different sources}
%\label{tbl-1}
\begin{center}\small
\begin{tabular}{lccccc}
Epoch & $\alpha_1$ & $E_{break,1}$ & $\alpha_2$ & $E_{break,2}$ &
$\alpha_3$\\
      &            & keV           &            & keV           &  \\
\hline
&&&&\\
Mkn~501--97 April 17$^a$ & 0.40$^{+0.02}_{-0.04}$ & 2.14$^{+0.3}_{-0.5}$ 
& 0.59$^{+0.02}_{-0.01}$ & $\sim 20$ & $0.84 \pm 0.04$ \\
Mkn~501--98 April 29 & $0.48 \pm 0.04$ & $1.3 \pm 0.4$ & $0.83 \pm 0.05$ & $17 \pm 5$ &
$1.15 \pm 0.05$\\
Mkn~501--98 May 01 & $0.62 \pm 0.04$ & $1.9 \pm 0.4$ & $1.00 \pm 0.05$ & $23 \pm 5$ &
$1.45 \pm 0.05$\\
&&&&\\
Mkn~421--97 April 21$^b$ & $0.85 \pm 0.03$ & $1.13 \pm 0.11$ & $1.40 \pm 0.1$ & $3.73 \pm 1.00$ & 
1.68$^{+0.12}_{-0.09}$\\
Mkn~421--98 May 4 & $1.30 \pm 0.03$ & $1.13 \pm 0.16$  & $1.67 \pm  0.4$ & $2.93 \pm 0.56$ 
& $2.15 \pm 0.1$\\
&&&&\\
PKS~2155--97 Nov$^b$ & $1.06 \pm 0.02$ & $1.09 \pm 0.07$ & $1.59 \pm 0.1$  
& $3.24 \pm 1.00$  & 1.88$^{+0.07}_{-0.03}$ \\  
%&&&&&\\
\hline
\multicolumn{6}{l}{$^a$ $\alpha _1$ and $\alpha _2$ are obtained from  broken 
power law fits of the LECS+MECS data;}\\
\multicolumn{6}{l}{\, \, $\alpha _3$ is obtained from broken power law fits
to  MECS+PDS data.}\\
\multicolumn{6}{l}{$^b$ $\alpha _1$ is obtained fitting  LECS data with a broken 
power law, 
$\alpha _3$ is obtained from}\\
\multicolumn{6}{l}{\, \, broken power law fits to  MECS data, while $\alpha _2$ 
is the average of the spectral indices}\\ 
\multicolumn{6}{l}{\, \, obtained from the previous fits in the common
anergy range of LECS and MECS data.}\\
\end{tabular}
\end{center}
\label{}
\end{table}

The 2-10 keV flux  observed from  Mkn~501  in April-May 1998 was
close to that measured on 7 April 1997, namely the lowest
 observed with \BS, but substantially  higher than
observed historically. In 1998
the synchrotron peak was located at an energy of $\sim$20 keV, much lower  than 
on April 97, but still exceptionally high compared to the  two sources discussed
above and even to most blue blazars, except possibly for
1ES2344+514 (Giommi et al. 1998).
As in the cases shown above, the synchrotron peak is at higher energies
 during brighter states and  in a given energy range the spectra flatten with
increasing intensity. However spectral variations are small at energies much 
below
the peak.

The TeV flux measured in April-May 1998 by the Whipple
and HEGRA telescopes was definitely lower than
observed last year at the beginning of the simultaneous X-ray and TeV
outburst. Since the X-ray spectra at the two epochs differed
mainly in the 20-100 keV band, we are led to conclude that
the TeV emission is most likely produced through IC
scattering off the high energy electrons which radiate in the
hard X-ray band via the synchrotron mechanism.
If this hypothesis is correct, then, analogously to the
synchrotron radiation peak, also the IC peak must have shifted toward
 lower energies causing significant spectral changes in the TeV band.

%%%%%%%%%%%%%%%%%%%%%%%%%%%%%%%%%%%%%%%%%%%%%%%%%%%%%
\section{PRESENT UNDERSTANDING AND OPEN PROBLEMS}
%%%%%%%%%%%%%%%%%%%%%%%%%%%%%%%%%%%%%%%%%%%%%%%%%%%%%%%%%

 The simplest model one can consider, 
applicable to all the sources discussed above, attributes the high
energy radiation to SSC emission from a
homogeneous spherical region of radius R, whose motion can be
characterized by a Doppler factor $\delta$, pervaded by a magnetic
field B and filled with relativistic particles with energy distribution 
described
by a bpl (the latter corresponds to 4 parameters: two
indices $n_1$, $n_2$, a break Lorentz factor $\gamma _b$ and a
normalization constant).

This seven parameter model is strongly constrained if observations provide
 a determination of the two slopes (in the X- and $\gamma$-ray
bands), the frequency and flux of the synchrotron peak, the frequency and flux
 of the IC peak. Assuming $R= c t_{var}$
% with a variability timescale $t_{var}=2$ hr,
the system is practically closed.  We refer to
Tavecchio, Maraschi \& Ghisellini (1998) for a general analytic 
procedure to determine  the physical parameters  of different sources
in this class of models. 
The main point we wish to stress is that there is little uncertainty on the 
model
parameters if both peaks can be measured simultaneously, and even when some of 
the
values (e.g., the peak energy of the IC component) are lacking,
the shape of the SED constrains the parameters in relatively restricted ranges.
In fact, different authors find closely similar parameters even when using
somewhat different formulations of the SSC model (e.g., 
Mastichiadis \& Kirk 1997, Ghisellini, Maraschi, Dondi 1996 for Mkn 421).

As an illustration, we show in Fig. 4 (left panel)
two SEDs computed  for PKS 2155--304 with the  SSC model described above
(the parameters are reported in the figure caption).
The scope was  to reproduce 
the lower and higher X-ray states of November 1997,
  together with the  $\gamma$-ray
data from the discovery observation (Vestrand, Stacy, \& Sreekumar
1995) and
the brighter $\gamma$-ray state of November 1997 (Sreekumar \&
Vestrand 1997).
 We (arbitrarily) assumed that the lower and higher X-ray intensity states
 correspond to the lower and higher $\gamma$-ray states, respectively.
In order to account for the flaring state, leaving the other
parameters unchanged, the break energy of the electron spectrum had to
be shifted to higher energies by a factor 1.5.  
As a consequence, both the synchrotron and IC peaks
increase in flux and move to higher energies. However, for the latter
the  effects are reduced with respect to the ``quadratic" relations expected in
the Thomson limit, since for these very high energy electrons the
suppression due to the Klein-Nishina regime plays an important role.

\footnote{The Compton emission is computed here with
the usual step approximation for the Klein-Nishina cross section,
i.e. $ \sigma = \sigma _T$ for $\gamma \nu _t < mc^2/h$ and $\sigma =0$ 
otherwise, where $\gamma $ is the Lorentz factor of the electron and 
$\nu _t$ is the frequency of the target photon.}

\begin{figure}
\epsfysize=8cm % fix the y-dimension and scales x-dim. to y-dim.
\vspace*{-1.0cm}
\centerline{ \hspace*{5.8cm} \epsfbox{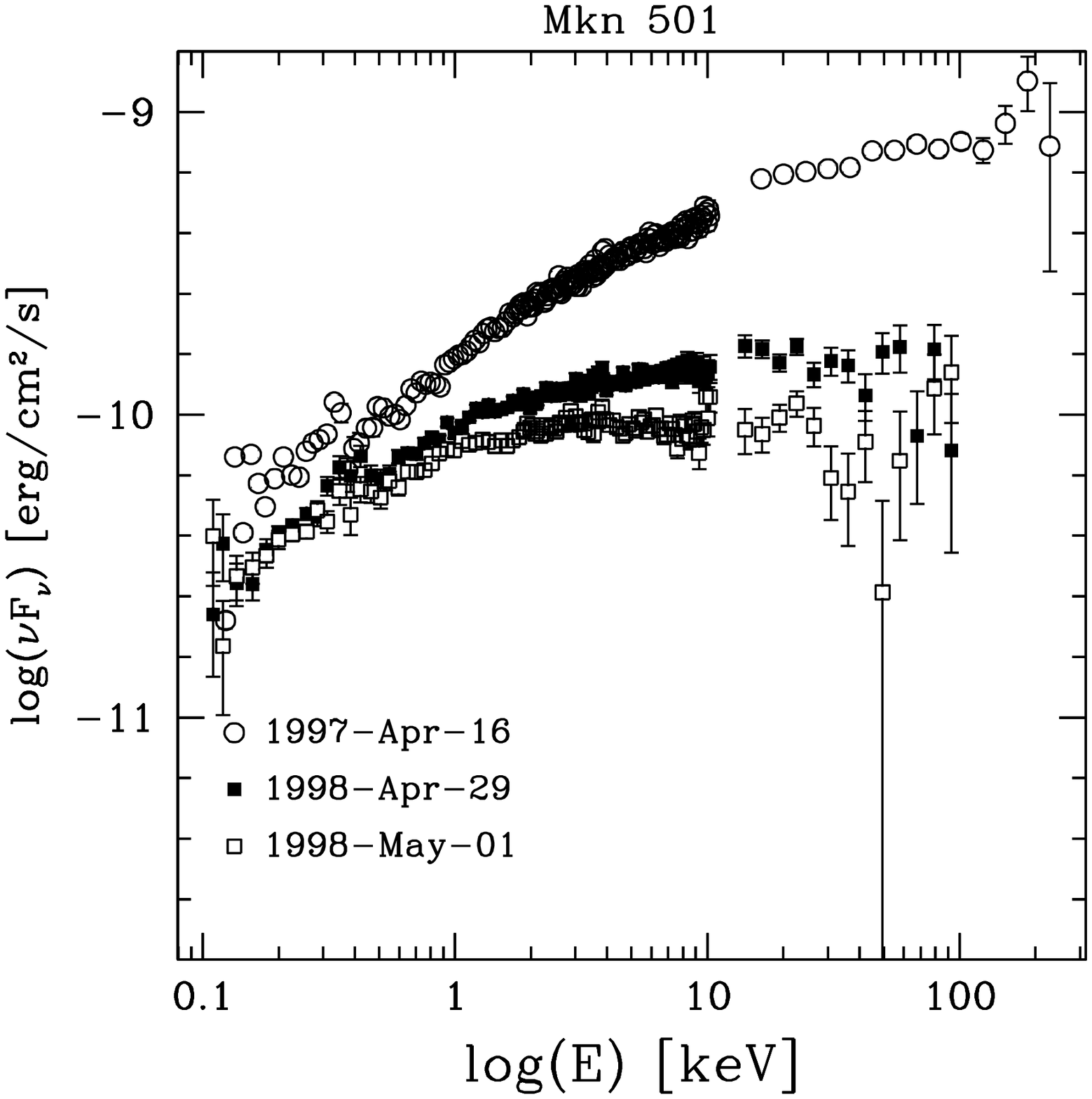} }
\epsfysize=8cm % fix the y-dimension and scales x-dim. to y-dim.
\vspace{-8.cm}
\hspace*{0.2cm}
\epsfbox{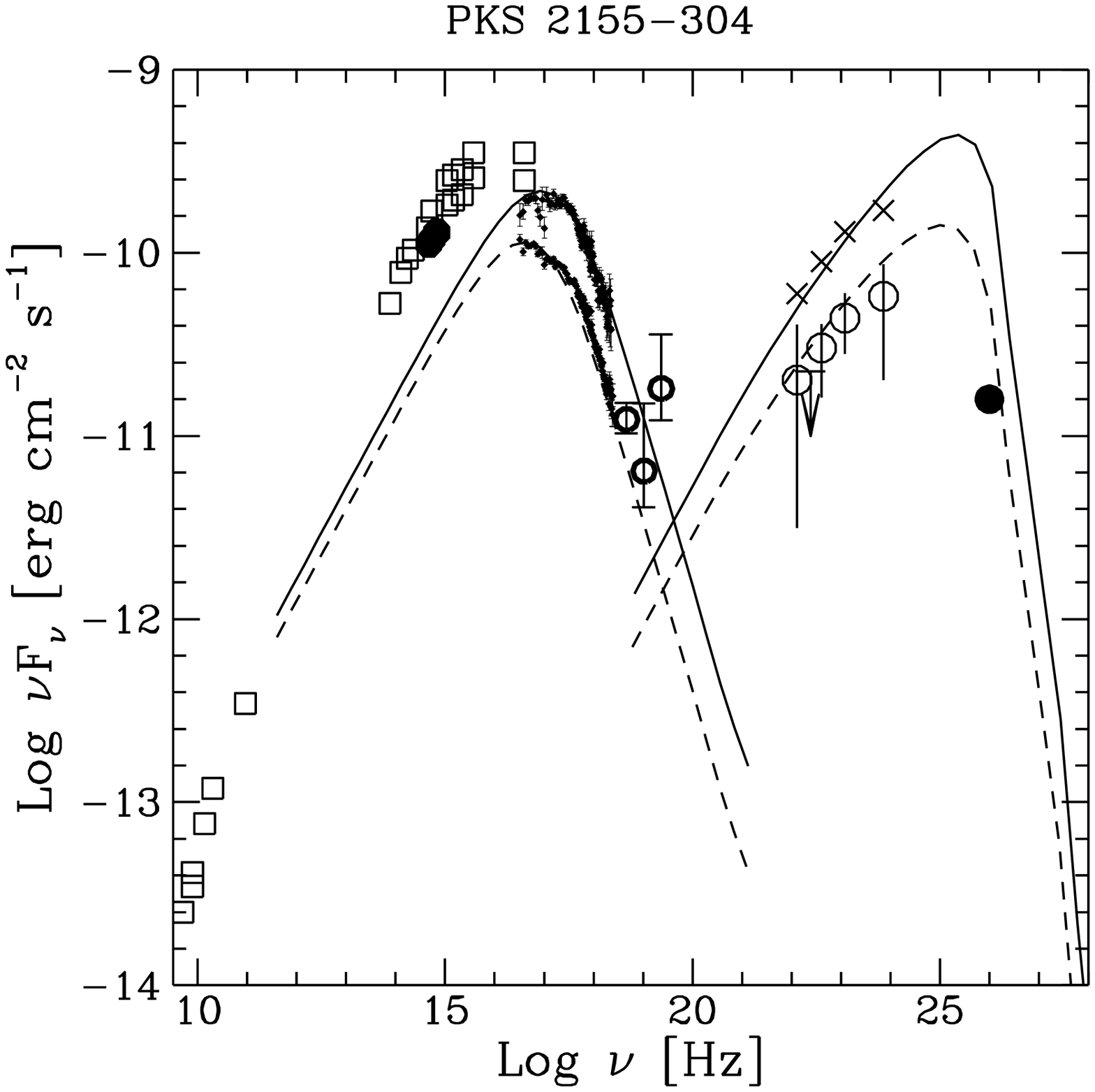}
\vspace*{0.2cm} 
\caption[h]{{\it Left panel}: SED of PKS 2155--304 compared with computed SSC 
models.
X-ray data points are from the high and low intensity states measured
by \BS in November 1997. The data at lower frequencies are from
previous non simultaneous observations. Open circles represent the $\gamma$-ray
data of 1995 and crosses the bright $\gamma$-ray state which triggered
the \BS observations (from Maraschi  et al. 1998).
The filled circle represents the TeV flux recently announced by Chadwick et al. 
(1998).
 The values of the parameters used in the model computations are: 
$\delta =18$, $B$=1 G, $K=10^{4.68}$, $R=3\cdot 10^{15}$ cm, 
$\gamma _b=10^{4.49}$, $n _1=2$, $n_2=4.85$ (low state) and $\delta =18$,
 $B$=1 G, $K=10^{4.8}$, $R=3\cdot 10^{15}$ cm, $\gamma _b=10^{4.65}$, $n _1=2$,
 $n_2=4.85$ (high state). {\it Right panel}: Deconvolved X-ray spectra of Mkn 
501 
obtained from \BS
 during  1997 April 16 (open circles), 1998 April 29 (filled squares) and 
 1998 May 1 (open squares).  
}
\end{figure}

The models predict TeV emission at a detectable level. Indeed, towards
the completion of this work, we have been informed of the detection of
high energy $\gamma$-rays by the Mark 6 telescope.
In November 1997 the source was seen at its highest flux (Chadwick et
al. 1998). 
The time averaged flux corresponds to $4.2\times 10^{-11}$ ph
 cm$^{-2}$ s$^{-1}$
above 300 GeV (and extending up to $>$ 3 TeV). 
In fact, the model for the lower intensity state
reproduces the TeV emission flux level remarkably well.
More observations are needed to study the correlation between the TeV
and X-ray fluxes in this source.

%%%%%%%%%%%%%%%%%%%%%%%%%%%%%%%%%%%%
\section{DISCUSSION AND CONCLUSIONS}
%%%%%%%%%%%%%%%%%%%%%%%%%%%%%%%%%%%%

The X-ray spectra and spectral variability of the three sources discussed above,
which are the brightest blue blazars in the X-ray band,
appear extremely coherent in the sense that they can be described
using analogous spectral laws, where differences between
different sources and different states of the same source can be
understood as changes in the energy at which the peak power is emitted.
Moreover their TeV emission is generally correlated with the X-ray intensity 
level within the same source.   

This behavior can be well understood in the context of the SSC process
in a relativistic jet, which seems quite convincingly to account for the
SED of blue  blazars. In these objects, of relatively low intrinsic power,
 the synchrotron and IC
components  tend to peak at the highest energies (X-ray and TeV
energies, respectively) and the synchrotron photons dominate the seed
radiation field upscattered to $\gamma$-ray energies.
The models allow us to establish the physical parameters of the jet plasma 
with relatively little uncertainty and there is general agreement that at least 
for
the three sources discussed above the required values of the Doppler
factors are quite high ($\delta \geq 10$), the magnetic fields are $10^{-1} - 1$ 
G
and the critical electron energy (the break energy in the bpl SSC model) are
of order $10^5$ with values as high as $10^6$ for Mkn 501 during the 1997 
high state.

While the assessment of the radiation mechanisms and physical parameters in the 
jet 
seems quite reliable, the inferred variations in the critical
electron energies, which seem to correlate well with brightness,
potentially contain  important clues for  an understanding of the 
variability and of the modes of particle acceleration and injection. 
Time dependent models for the evolution of the energy distributions of electrons
subject to acceleration, injection, propagation and energy losses are required. 
  The problem is in general complex and
only some simplified cases have been treated up to now (e.g., Kirk et
al. 1998, Dermer et al. 1998, Makino 1998). In addition,
light travel time effects through the emitting region may be important
(Chiaberge \& Ghisellini 1998).

An important point  is the measurement and interpretation
of lags of the soft photons with respect to the harder ones. These can
be due to radiative cooling if the population of
injected (accelerated) electrons has a low energy cut off or possibly
a sharp low energy break, as clearly shown by Kazanas et al. (1998).
If so, the observed lag $\tau _{obs}$ depends only on the
value of the magnetic field (assuming synchrotron losses are dominant, 
as is roughly the case for these sources) and on $\delta$. Their relation 
can be expressed as
\begin{equation}
B\delta^{1/3}=300 \left( \frac{1+z}{\nu _{1}} \right)^{1/3}  
\left[ \frac{1-(\nu_1/ \nu_0)^{1/2}}{\tau _{obs}} \right]^{2/3} G
\label{diciassette}
\end{equation}
where $\nu_1$ and $\nu_0$ represent the frequencies (in units of 
$10^{17}$ Hz) at which the
observed lag has been measured.

It is interesting to note that the value of the soft lag inferred for PKS 
2155--304
from the \BS observations yields a B and $\delta$ combination consistent
with the parameters obtained {\it independently} from the spectral
fitting, which supports the
radiative interpretation of the observed X-ray variability (Tavecchio, Maraschi, 
\&
Ghisellini 1998). 
However the soft {\it lead} possibly present in the 1998 flare of Mkn 421
should probably be related to the acceleration time scale (Kirk et al. 1998), 
implying
that  the time dependence of acceleration/injection events themselves plays 
a significant role.

In conclusion,
the time resolved continuum spectroscopy, made possible by sensitive
and broad band instruments like ASCA and \BS, has allowed us to
reconstruct
the spectra of the emitting high energy electrons in blazars and to follow 
their evolution in time. Therefore X-ray data of the present quality 
allow us to probe not only the radiation mechanisms but also the fundamental
 processes of particle acceleration and transport in relativistic jets.

\section{REFERENCES}

Aharonian, F., et al. 1997, A\&A, 327, L5 \\
Bicknell, G.V., 1994, ApJ, 422, 542\\
Blandford, R.D., \& Rees, M.J., 1978, in Pittsburgh Conf. on BL Lac
objects, ed. AM Wolfe, 328\\
Catanese, M., et al. 1997, ApJ, 487, L143 \\
Chadwick, P.M., et al., 1998, ApJ, in press (astro-ph/9810209) \\
Chiaberge, M., \& Ghisellini, G., 1998, MNRAS, submitted (astro-ph/9810263)\\
Chiappetti, L. \& Torroni, V. 1997, IAU Circ., 6776, 2\\
Chiappetti, L., et al., 1998, submitted\\
Dermer, C. D., Sturner, S. J., \& Schlickeiser, R. 1997, ApJS, 109, 103\\
Dermer, C. D. 1998, ApJ, 501, L157\\
Edelson, R. A., \& Krolik, J. H. 1988, ApJ, 333, 646\\
Fossati, G., et al., 1997, MNRAS, 289, 136\\
Fossati, G., et al., 1998, MNRAS, 299, 433\\
Ghisellini, G., Maraschi, L., Dondi, L., 1996, A\&AS, 120, 503\\
Ghisellini, G. \& Maraschi, L., 1996, ASP Conf. Ser., 110, 436\\
Ghisellini, G., et al., 1998, Nucl. Phys. B Proc. Suppl., 69, 427\\
Giommi, P., et al., 1998a, Nucl. Phys. B Proc. Suppl., 69, 407\\
Giommi, P., et al., 1998b, A\&A, 333,L5\\
Kazanas, D., Titarchuk, L. G., \& Hua, X.-M. 1998, ApJ, 493, 708\\
Kirk, J.G., Rieger, F.M., \& Mastichiadis, A. 1998, A\&A, 333, 452\\
Lockman, F. J. \& Savage, B. D. 1995, ApJS , 97, 1\\
Mastichiadis, A. \& Kirk, J.G., 1997, A\&A, 320, 19\\
Makino, F., 1998, to be published in BL Lac phenomenon\\
Maraschi, L., et al., 1994, ApJ, 435, L91\\
Maraschi, L., 1998, Nucl. Phys. B Proc. Suppl., 69, 389\\
Maraschi, L., et al., 1998, to be published in "Tutti i colori degli AGN", 
third italian conference on AGN, Roma May 18-21, Memorie SAIt 
(astro-ph/9808177)\\
Padovani, P., et al., 1998, Proc. of the Conference ``BL Lac Phenonmenon" 
(Turku,
Finland), PASP Conf. Ser., eds. L. Takalo, in press\\
Pian, E., et al. 1998, ApJ, 492, L17\\
Sikora, M. 1994, ApJS, 90, 923\\
Sreekumar, P. \& Vestrand, W. T. 1997, IAU Circ., 6774, 2\\
Takahashi, T., et al., 1996, ApJ, 470, L89\\
Tavecchio, F., Maraschi, L., Ghisellini, G., 1998, ApJ, in press 
(astro-ph/9809051)\\
Treves, A., et al., 1998, Proc. of the Conference ``BL Lac Phenonmenon" (Turku,
Finland), PASP Conf. Ser., eds. L. Takalo, in press (astro-ph/9811244)\\
Ulrich, M.H., Maraschi, L., Urry, C.M., 1997, Ann. Rev. Astron. and Astroph., 
35,
445\\
Urry, C. M., et al. 1997, ApJ, 486, 799\\
Vestrand, W. T., Stacy, J. G., \& Sreekumar, P. 1995, ApJ, 454, L93\\
Wehrle, A. E., et al. 1998, ApJ, 497, 178\\
Wolter, A., et al., 1998, A\&A, 335, 899\\

\end{document}